\documentclass[pre,twocolumn, a4paper, superscriptaddress,
showpacs, amsmath, nofootinbib, floatfix] {revtex4}

\usepackage[mathcal]{eucal}
\usepackage{bm}
\usepackage{graphicx}

\setkeys{Gin}{width=1\linewidth}
 
\begin{document}

\title{Roughness distributions for $1/f^{\alpha}$ signals}


\author{T. Antal} \email{Tibor.Antal@physics.unige.ch}
\affiliation{ D\'epartement de Physique Th\'eorique,
  Universit\'e de Gen\`eve, CH 1211 Gen\`eve 4, Switzerland}
\affiliation{Institute for Theoretical Physics, E\"otv\"os University,
  1117 Budapest, P\'azm\'any s\'et\'any 1/a, Hungary}

\author{M. Droz}
\email{Michel.Droz@physics.unige.ch}
 \affiliation{ D\'epartement de Physique Th\'eorique,
  Universit\'e de Gen\`eve, CH 1211 Gen\`eve 4, Switzerland}

\author{G. Gy\"orgyi}
\email{gyorgyi@glu.elte.hu}
\affiliation{Institute for Theoretical Physics, E\"otv\"os University,
  1117 Budapest, P\'azm\'any s\'et\'any 1/a, Hungary}

\author{Z. R\'acz} 
\email{racz@poe.elte.hu} \affiliation{Institute for
  Theoretical Physics, E\"otv\"os University, 1117 Budapest,
  P\'azm\'any s\'et\'any 1/a, Hungary} \affiliation{ Laboratoire de
  Physique Th\'eorique, B\^atiment 210, Universit\'e de Paris-Sud,
  91405 Orsay Cedex, France}

\date{\today}


\begin{abstract}
  
  The probability density function (PDF) of the roughness, i.e., of
  the temporal variance, of $1/f^\alpha$ noise signals is studied.
  Our starting point is the generalization of the model of Gaussian,
  time-periodic, $1/f$ noise, discussed in our recent Letter
  \cite{us}, to arbitrary power law.  We investigate three main
  scaling regions ($\alpha \leq 1/2$, $1/2<\alpha\leq 1$, and
  $1<\alpha$), distinguished by the scaling of the cumulants
  in terms of the microscopic scale and the total length of the
  period.  Various analytical representations of the PDF allow for a
  precise numerical evaluation of the scaling function of the PDF for
  any $\alpha$.  A simulation of the periodic process makes it
  possible to study also non-periodic, thus experimentally more
  relevant, signals on relatively short intervals embedded in the full
  period.  We find that for $\alpha\le 1/2$ the scaled PDF-s in both
  the periodic and the non-periodic cases are Gaussian, but for
  $\alpha>1/2$ they differ from the Gaussian and from each other.  Both
  deviations increase with growing $\alpha$.  That conclusion, based
  on numerics, is reinforced by analytic results for $\alpha=2$ and
  $\alpha\to \infty$, in the latter limit the scaling function of the
  PDF being finite for periodic signals, but developing a singularity
  for the aperiodic ones.  Finally, an overview is given for the
  scaling of cumulants of the roughness and the various scaling
  regions in arbitrary dimensions. We suggest that our theoretical and
  numerical results open a new perspective on the data analysis of
  $1/f^\alpha$ processes.

\end{abstract}

\maketitle


\section{Introduction}
\label{sec:intro}

The power spectra of fluctuations scales with frequency as
$S(f)\sim1/f^{\alpha}$ in a large variety of physical, chemical and
biological systems \cite{hong-kong}.  This power law behavior
$1/f^{\alpha}$ often persists over several orders of magnitude with
cutoffs present at both high and low frequencies, and with typical
values of $\alpha$ in the range $0.8 \le \alpha \le 4$
\cite{hong-kong}.  In a somewhat loose terminology, all these systems
are said to display $1/f$ noise although good quality data with
$\alpha$ very close to $1$ exist only for the voltage fluctuations
when a current is flowing through a resistor \cite{Weissman,Yakimov}.
Phenomena with $\alpha\not=1$, however, are abundant, examples being
the white-dwarf light emission \cite{press}, the flow of sand through
hourglass \cite{shick}, ionic current fluctuations in membrane
channels \cite{siwy}, number of daily trades in the stock market
\cite{mantegna}, water flows of rivers \cite{rivers}, the spike trains
of nerve cells \cite{cells}, the occurrence of earthquakes
\cite{earthquakes}, the traffic flow on a highway
\cite{highway1,highway2}, the electric noise in carbon nanotubes
\cite{nano} and in nanoparticle films \cite{nanofilms}, the interface
fluctuations \cite{Krugreview}, dissipation in turbulent systems
\cite{turbulence}, and the list could be continued.

A well understood example of $1/f^{\alpha}$ type behavior is the
dynamic scaling observed at equilibrium critical points where the
power-law correlations in
time are generated by the infinite-range correlations in space.
Most of the examples listed above, however, are related to non-equilibrium
phenomena and a similar level of understanding does not exist.
There have been many attempts at identifying possible
generic mechanisms leading to scale invariant fluctuations, a notable
example being the concept of self organized criticality
\cite{BTW,bak}. It is clear, however, that not all the systems
showing a $1/f^{\alpha} $ noise fit into a single scheme and, perhaps,
at this stage one should pursue a less ambitious aim of developing
more detailed characterizations of non-equilibrium universality
classes.

In equilibrium systems, the static universality classes are determined
by the (i) dimensionality, (ii) symmetry of the order parameter, and
(iii) the range of the interactions. Further specifying the
conservation laws and the coupling of the order parameter to conserved
quantities defines then the dynamical universality classes. The
statics and dynamics are, however, intertwined in non-equilibrium
systems and an exponent in the $1/f^{\alpha}$ behavior carries
information about both. Thus measuring a single exponent $\alpha$ (a
characteristic situation when measuring time-series of noise) does not
determine the universality class of the system even within the
framework of an equilibrium-type theory. Having a single exponent,
however, one can still go on and try to ascertain whether two systems
belong or not to the same universality class. This can be done by
using the data to measure and compare the scaling functions associated
with the finite-size scaling of some global physical quantity such as,
e.g., the order parameter in a critical system or the roughness of an
interface.

The remarkable features of scaling functions are that they are
obtained without any fitting procedure and, furthermore, they usually
converge fast as the system size is increased. Thus one can build a
picture gallery that can be effectively used to identify systems
belonging to a given universality class. Indeed, such an approach has
been useful in establishing connections among rather diverse processes
such as massively parallel algorithms \cite{Korniss}, interface
dynamics in the $d=2$ Fisher-Kolmogorov equation \cite{Tripathy},
dissipation fluctuations in a turbulence experiment
\cite{Pinton1,Pinton2} and the interface fluctuations in the $d=2$
Edwards-Wilkinson model ($XY$ model) \cite{Holdsworth1,Holdsworth2}.
This scaling function approach has also helped to clear up some
questions about the upper critical dimension of the
Kardar-Parisi-Zhang equation \cite{KPZ}.
 
In our attempts to expand the picture gallery, we have recently
derived \cite{us} the following result: The scaling function for the
roughness-distribution of a Gaussian periodic $1/f$ noise signal is
one of the extreme value distributions, the Fisher-Tippet-Gumbel
distribution \cite{Gumbel,Galambos}. This is a rather unexpected and
interesting result and we feel that it is important to explore its
generality and limitations.  First, because ideas about extreme
statistics playing a role in strongly correlated, scale invariant
systems have been much discussed
\cite{Holdsworth1,Holdsworth2,Jensen,Doussal, RCPS}, and this result may
provide a foundation to those speculations.  Second, because this is
the first instance when one of the distributions associated with
extreme statistics emerges naturally, and in a mathematically precise
manner, for a quantity which is not of extremal character a priori.

There are three elements underlying the above extreme statistics
results: Gaussianity, $1/f$ power spectrum, and periodicity. Thus the
problem can be generalized by considering effects of (1)
non-gaussianity, (2) a generalized power spectra of the form
$1/f^\alpha$, and (3) non-periodicity (the experimentally realistic
situation).  In this paper, we shall concentrate on points (2) and (3)
leaving the more difficult problem of non-gaussianity to a later
study. 
 
One expects that changing the exponent $\alpha$ will change the
roughness distribution of a signal and, indeed, there are analytical
results \cite{{us},{FORWZ},{PRZ94}} for $\alpha=1,2,4$ which
demonstrate this explicitly for periodic signals (d=1 dimensional
systems). Similar results exist also for higher dimensions where the
roughness of a d=1 signal is replaced by the roughness of a higher
dimensional interface \cite{RP94,Holdsworth1,Holdsworth2,Duna}.  Thus
the questions we address in connection with the $\alpha$-dependence is
not about its existence but about its magnitude (observability).  More
specifically, we ask if there is an interval of $\alpha$ where the
$\alpha$-dependence is absent or negligible from experimental point
of view? 
 
The other problem of our concern is the question of the effect of the
boundary conditions (BC-s). When analyzing a signal and building a
distribution function, one usually divides the signal into equal
"time" intervals and measures the quantity of interest in every
interval. Thus the signal within a given interval is not periodic
(usually the total signal is not periodic either) and the question
arises whether the experimentally used BC, i.e., when one studies a
"window" within the signal, would affect the scaling functions or not.
This shall be a central issue in this paper, so we introduce the
abbreviations PBC and WBC, for periodic and ``window'' BC,
respectively. 
 
The dependence of scaling functions on the boundary conditions has
been discussed in equilibrium critical phenomena and it is known that
such dependence exist \cite{Wu,exotic}. Thus our aim here again is not
to show its presence but to gauge its magnitude for various $\alpha$
values and investigate for which values of $\alpha$ this effect is
absent or negligible.
 
As we show below, the scaling functions can be derived analytically in
case of PBC. Similar analytical treatment for WBC were achieved only
for $\alpha=2$ and in the $\alpha\rightarrow\infty$ limit, otherwise
we had to resort to simulations in order to obtain the finite alpha
results. Observing the scaling functions (displayed on the Figures
below) one can deduce the following qualitative trends as $\alpha$ is
decreased.

(1) The PBC scaling functions change smoothly apart from
$\alpha\approx 1$ where the function is sensitive to small changes in
$\alpha$ since a singularity develops at $\alpha=1$. Even this
singular behavior can be smoothed out, however, by an appropriate
change of the scaling variable. The scaling function approaches a
Gaussian as $\alpha\rightarrow 1/2$ and it remains a Gaussian for
$\alpha<1/2$. 

(2) The WBC scaling functions display strong $\alpha$ dependence at
large $\alpha$. The dependence becomes weaker for $\alpha\alt 2$ and,
similarly to the PBC case, the $\alpha$-dependence disappears entirely
for $\alpha\le 1/2$ where the function becomes a Gaussian. 

(3) Comparing the PBC and WBC scaling functions, one can observe that
their difference is large for $\alpha\agt 4$, it is easily noticeable
in the range between $1$ and $4$, while it becomes harder to
distinguish the functions for $\alpha\alt 1$, and the
functions become identical Gaussians for $\alpha\le 1/2$.

In order to demonstrate the above observations, we shall introduce a
model of $1/f^\alpha$ noise in Sec.\ \ref{sec:roughness} where the
roughness distribution, the appropriate choice of scaling variables,
and some general scaling properties will also be discussed. Sec.\ 
\ref{sec:scaling-fcn-pbc} contains the analytical calculations for the
case of PBC while Sec.\ \ref{sec:scaling-fcn-wbc} is devoted to
presenting both the simulation and analytical results for the WBC
case. We close with discussing generalizations to higher dimensions in
Sec.\ \ref{sec:gen-dim}.

\section{Probability distribution of periodic $1/f^{\alpha}$ signals
  and their roughness}   
\label{sec:roughness}
 
\subsection{Model for periodic, perfectly $1/f^\alpha$, Gaussian  noise}
\label{ssec:model}

Noisy signals are in principle fully characterized by their path
probability density functional.  However, the most often used
characterization is by the power spectrum, whose experimental
recording is straightforward.  This leaves several properties
unspecified, such as the distribution of the phases, and whether the
noise is Gaussian.  Furthermore, the usual Fourier expansion of the
signal in an interval implies periodicity outside the interval.  While
that expansion can be used well for data analysis, the physical
implication for the outer part of the signal, which influences by
long-time correlations the inner part, is obviously non-realistic.
Accordingly, the role of initial and final conditions in time, a
property that we shall call boundary condition (BC), are generically
neglected.  Whereas it is arguable that in a stationary process, long
time after the turning on of the experimental device, the boundary
conditions should not have much influence, when a model is constructed
for $1/f^\alpha$ noise, the boundary conditions should be defined with
care.

We now describe the main properties of the model treated in this
paper.  It should be emphasized that we do not study the underlying
microscopic mechanism that may lead to $1/f^\alpha$ noise, this has
been done on many different levels of simplification for a large
number of systems, for a review see \cite{Weissman}.  Rather, we
construct a stochastic model, where the noisy signal is given by its
path probability density functional, having the generic properties of
observed $1/f^\alpha$ noises.  In a sense it will be a minimal model
tailored to exhibit a few properties we prescribe.

Firstly, the noise we consider is periodic.  While this is  obviously
not valid in a real experimental situation over the time interval of
the measurement, as we have seen recently \cite{us}, it yields
theoretical predictions not very far from what is measured in a real
signal.  Furthermore, the periodic boundary condition (PBC) allows for
the straightforward numerical simulation of the noise, thus making it
possible to study numerically the statistics in cut-out time intervals
that are much smaller than the entire period.  Such a ``window''
boundary condition (WBC) is, we surmise, a more faithful
representation of experimentally realistic BC-s.  Indeed, while the
effect of the outer signal onto the inner part is important due to
long{\-}\hbox{-time} correlations, probably it depends little on
whether the signal was simply turned on long time ago, or is periodic
on a time scale much larger than the ``window''.  Looking at it from
another angle we can say that the PBC is equivalent to saying that the
signal is best expanded in the usual Fourier basis, thus what remains
is to specify the probability distribution of the Fourier
coefficients.  Note that the noisy signal in a ``window'' should not
be expanded in such a basis in theory, because the noise is manifestly
aperiodic.

Secondly, we restrict ourself to Gaussian noise and assume that the
Fourier modes are independent random variables. The reasons for this
working hypothesis are that many experimental data concern power
spectrum measurements of Gaussian noises and provide no information
about the coupling between the modes.

Thirdly, we assume that the phase of each complex Fourier component is
random and uniformly distributed in $[0,2\pi]$.  Hence the probability
distribution will depend only on the modulus. 

Last but not least, we consider perfectly $1/f^\alpha$ noise.  Here we
understand that the variance of the independent Fourier amplitudes of
a path is a pure power with exponent $\alpha$.  Furthermore, there are
no upper and lower cutoff frequencies other than the natural ones
determined by the observation time and the microscopic characteristic
time unit, respectively. 
 
The above ingredients specify the path probability of the stochastic
time signal, this will be our model, as discussed in detail below.
(Note that it is a generalization of the case $\alpha=1$ \cite{us} to
arbitrary $\alpha$.)

The stochastic trajectories $h(t)$ are periodic, $h(t)=h(t+T)$, thus
the trajectory can be expanded as
\begin{equation}  
  \label{eq:fourier} 
  h(t) = \sum_{n=-N}^{N} c_n\, e^{2\pi int/T}, \quad c_n^\ast = 
  c_{-n}. 
\end{equation} 
Here we have used a cutoff $N$ meaning that the time scale is not
resolved below $\tau =T/N$.  This is needed to deal with some
singularities and may have microscopic physical origin.  Whenever
possible we shall, however, take $N\to\infty$, and even if this were a
singular limit, we shall wind up with scaling functions not containing
the microscopic time unit, so the value of $\tau$ may be left
undetermined. 

The stochastic properties of a signal $h(t)$ are fully characterized
by our specifying the probability density functional
\begin{equation}
  \label{eq:path-pdf}
  P[h(t)] = e^{-S[h(t)]},
\end{equation}
or, equivalently, the probability density function (PDF) for the
Fourier coefficients $c_n$ for $n=0\dots N$ as 
\begin{equation}
  \label{eq:path-pdf-fourier}
 P\left( \left\{ c_n \right\} \right) = A\, e^{-S\left(
 \left\{ c_n \right\} \right) }.
\end{equation}
Our model for periodic, perfectly $1/f^\alpha$ signals is defined by the
action
\begin{eqnarray}
  \label{eq:def-action}
 S\left(
 \left\{ c_n \right\}\right) &=& 2\sigma\, T^{1-\alpha}
 \sum_{n=1}^N n^\alpha\, \left| c_n\right| ^2, \\
 A &=& \prod_{n=1}^N 2 \sigma n^\alpha T^{1-\alpha}\pi^{-1},
\end{eqnarray}
where the probabilistic variables are the real and imaginary parts of
$c_n$-s.\footnote{In time-representation the action is of the form
\begin{equation}
  \label{eq:def-action-time}
 S[h(t)] = 2\sigma \left(\frac{-1}{2\pi}\right)^\alpha
 \int_0^T dt\, \left( \frac{d^{\alpha/2} h}{dt^{\alpha/2}}\right)^2,
\nonumber \end{equation}
where the fractional derivative is understood on the Fourier
representation so that it acts upon the phase factor as $d^{\beta}
e^{iat}/dt^{\beta} = (ia)^{\beta}e^{iat}$.  Dimensional analysis on
Eq.\ (\ref{eq:def-action-time}) yields the scaling by $T$ used in
(\ref{eq:def-action}).} We assume translation invariance in $h$ space,
therefore the action must not depend on the constant part $c_0$ and we
set that to zero hereafter.  Note furthermore that the above action
means that different Fourier modes are uncorrelated.  The coefficient
$\sigma$ makes $S$ dimensionless and can be understood as the
reciprocal noise intensity parameter.  Since we are after scaling
functions, the value of $\sigma$ will not matter. The above action was
set up so that the power spectrum is
\begin{equation}
  \label{eq:power-spectrum}
\left< \left| c_n\right|^2 \right> \propto
\frac{1}{n^\alpha}. 
\end{equation}
thus the process is indeed of the $1/f^\alpha$-type.

\subsection{Roughness of a signal}
\label{ssec:roughness} 
 
Our aim is to characterize the signals by some global properties. As
the time average $c_0$ in Eq.\ (\ref{eq:fourier}) was set to zero, the
simplest quantity worth considering is the mean square width, i.e.,
roughness of the signal.   This has been studied for the $1/f$ noise
($\alpha=1$) in Ref.~\cite{us}, the Wiener process ($\alpha=2$) in
Ref.~\cite{FORWZ}, and for curvature-driven interfaces ($\alpha=4$) in
Ref.~\cite{PRZ94}.  Its general definition is
\begin{equation}
  \label{eq:def-roughness}
  w_2 = \overline{ \left[ h(t) - \overline{h(t)}\right]^2 } =
  \overline{  h^2(t)} - \overline{h(t)}^2,
\end{equation}
where overbar means time average over the entire period $T$.  Note
that now $\overline{h(t)}=0$.  In Fourier representation we get
\begin{equation}
  \label{eq:def-roughness-fourier}
  w_2 = 2 \sum_{n=1}^N \left| c_n \right|^2,
\end{equation}
hence $w_2$ is in fact the integrated power spectrum. 

The roughness $w_2$ is associated with a given $h(t)$ trajectory, it
varies when different instances of the paths are taken from their
ensemble.  Thus $w_2$ is a probabilistic variable, whose PDF can be
expressed as
\begin{equation}
  \label{eq:pdf-roughness-time}
  P\left(w_2\right) = \int \delta\left(w_2 -
  \overline{  h^2(t)} + \overline{h(t)}^2 \right)\, P[h(t)]  {\mathcal
  D}h(t).
\end{equation}
In the more practical Fourier representation $P\left(w_2\right)$
assumes the form
\begin{eqnarray}
  P \left(w_2\right) &=& \int \delta\left(w_2 -
  2\sum_{n=1}^N \left| c_n \right| ^2 \right)\, P\left( \left\{
  c_n \right\}  \right)  \nonumber \\
  && \times\prod_{n=1}^N
  d\textrm{Re}c_n\,d\textrm{Im} c_n     \label{eq:pdf-roughness-fourier}
\end{eqnarray} 
It is useful to introduce the generating function of the moments
of $P\left(w_2\right)$ as
\begin{equation}
  \label{eq:gen-fun}
  G (s) = \int_0^\infty \!\! dw_2\, P (w_2)\, e^{-sw_2},
\end{equation} 
which by (\ref{eq:pdf-roughness-fourier}) gives
\begin{equation}
  \label{eq:gen-fun-roughness}
  G (s) = \prod_{n=1}^N \left( 1+\frac{s}{\sigma
  T^{1-\alpha} n^\alpha} 
  \right)^{-1}.  
\end{equation}
The cumulant generating function $\Psi (s)$ can be
obtained by expanding the logarithm as
\begin{equation}
  \label{eq:cum-gen-fun-roughness}
  \Psi (s)= \ln G (s) = \sum_{k=1}^\infty
  \frac{1}{k} \left( \frac{-s}{ 
  \sigma T^{1-\alpha}}  \right) ^k \sum_{n=1}^N \frac{1}{n^{ \alpha k}}.  
\end{equation}
Hence the $k$-th cumulant of the roughness is 
\begin{equation}
  \label{eq:cum-roughness}
  \kappa_k = \frac{(k-1)!}{\left(\sigma T^{1-\alpha}\right)^k}  \sum_{n=1}^N
  \frac{1}{n^{ \alpha k}}.   
\end{equation}
Note that for $\alpha k>1$ the cumulants can be expressed in terms of
Riemann's zeta function in the limit of large $N$ as
\begin{equation}
  \label{eq:cum-roughness-zeta}
  \kappa_k \to \frac{(k-1)!}{ \left(\sigma T^{1-\alpha}\right)^k}
  \zeta(\alpha k).   
\end{equation}
The low-order cumulants have the usual meaning of average and
variance 
\begin{subequations}
  \label{eq:low-cum-roughness}
\begin{eqnarray}
  \kappa_1&=&\left< w_2 \right>   = \frac{1}{ \sigma T^{1-\alpha}}
  \sum_{n=1}^N \frac{1}{n^{ \alpha}}, \label{eq:cum1-roughness} \\
  \kappa_2   &=& \left< w_2^2 \right> -\left< w_2 \right>^2  =
  \frac{1}{ \left(\sigma T^{1-\alpha}\right)^2} 
  \sum_{n=1}^N \frac{1}{n^{ 2 \alpha}}. \label{eq:cum2-roughness} 
\end{eqnarray} 
\end{subequations}

The PDF corresponding to the generator (\ref{eq:gen-fun-roughness}) will
be the main concern of this paper.  We shall also discuss ``window''
boundary condition, WBC, so in case of ambiguity the above quantities,
pertaining to PBC, will get a subscript like $G_{\textrm{p}}(s)$.
 
\subsection{Scaling of averages} 
\label{ssec:scaling} 
 
The generating function of the PDF of the roughness still contains two
scales, the observation time $T$ (multiplied by
$\sigma^{1/(1-\alpha)}$) and the microscopic time unit $\tau=T/N$.  If
$N$ diverges, physical quantities will exhibit various scaling
for different $\alpha$-s.  Whereas the above model is a very simple
one from the viewpoint of the statistical mechanics of critical
phenomena, as it is Gaussian and massless, it is worth summarizing the
scaling properties of averaged quantities, because they contain
important information about the PDF. 
 
The $k$-th roots of the $k$-th cumulants (\ref{eq:cum-roughness}) for all
$k$-s have the dimensionality of the roughness, however, their scaling in
the length $T$ of the time interval may be different.  For $k=1,2$
these are the average and the mean deviation therefrom, respectively.
Firstly, for $\alpha > 1$ all cumulants converge in the limit
$N\to\infty$ and we have 
\begin{equation} 
  \label{eq:cum-scale-large-alpha}  
  \sqrt[k]{\kappa_k} \propto T^{\alpha-1}.  
\end{equation}  
If $\alpha = 1$ then the average, $\kappa_1$, logarithmically diverges
in $N=T/\tau$ while all higher cumulants remain of unit order,
independent of $T$ and $\tau$.  This corresponds to a critical
dimension in second order phase transitions, but here we have a
critical $\alpha=1$ where the logarithmic singularity occurs.  For
$\alpha < 1$ the average no longer depends on $T$, rather it is
proportional to $\tau^{1-\alpha}$, and for $1/2 < \alpha < 1$ the
$k>1$ cumulants still exhibit the scaling
(\ref{eq:cum-scale-large-alpha}).  The second critical value of
$\alpha$ is at $1/2$, where the second cumulant has a logarithmic
singularity.  For $\alpha<1/2$ the $\sqrt[2]{\kappa_2}$ becomes
proportional to $T^{-1/2} \tau^{\alpha-1/2}$, while
higher-than{\-}{-second} order cumulants still satisfy
(\ref{eq:cum-scale-large-alpha}) until the third critical value of
$\alpha=1/3$ is reached.  Continuing the reasoning shows that at
$\alpha_k=1/k$ the $k$-th order cumulant develops a logarithmic
singularity.  A significant change occurs at $\alpha_2=1/2$, because
for lesser $\alpha$-s the scale of the mean deviation ($k=2$) will
dominate over the scale from higher $k$-s.  This means that the PDF of
the roughness on the scale of its mean deviation becomes Gaussian in
the $N\to\infty$ limit.  Thereby the effect of the sequence of
critical $\alpha_k$-s for $k>2$ is suppressed.  Table
\ref{tab:scaling} summarizes the scaling of the cumulants.  Note that
the powers of $T$ and $\tau$ add up to $\alpha -1$ for each entry to
produce the right time-dimensionality.

\begin{table}
\caption{\label{tab:scaling} Scaling of the cumulants for various
  $\alpha$-s.  The dots indicate that the last formula is valid
  from then on, i.e., ``$\dots$'' extends the validity to
  higher $k$-s and vertical dots to lower $\alpha$-s. } 
\begin{ruledtabular}
\begin{tabular}{cccccc}
Range or& & & & &\\
value of $\alpha$.  &$\kappa_1$ &$\sqrt{\kappa_2}$
&$\sqrt[3]{\kappa_3}$ &$\sqrt[4]{\kappa_4}$
& \\ 
\hline
$(1,\infty) $ & $T^{\alpha-1}$ & $\dots$ & & & \\
$1$ & $\ln\frac{T}{\tau}$ & O($1$) &$\dots$ &&\\
$(1/2,1)$ & $\tau^{\alpha-1}$ & $T^{\alpha-1}$ &$\dots$ &&\\
$1/2$ & $\vdots$ & $T^{-1/2}\sqrt{\ln\frac{T}{\tau}}$
& $T^{-1/2}$ &$\dots$ &\\
$ (1/3,1/2)$ &  &
$\tau^{\alpha-1/2}\, T^{-1/2} $ &$T^{\alpha-1}$ &$\dots$ &\\ 
$1/3$ &  &
$\vdots $ &
$T^{-2/3}\sqrt[3]{\ln\frac{T}{\tau}}$ & $T^{-2/3}$ &$\dots$ \\ 
$(1/4,1/3)$ & 
&  & $\tau^{\alpha-1/3}\,T^{-2/3}$ &$T^{\alpha-1}$&$\dots$\\ 
 & & &$\vdots$&&\\
\end{tabular}
\end{ruledtabular}
\end{table}

As mentioned above and can be seen in Table \ref{tab:scaling}, for
$\alpha<1/2$ the mean deviation goes with the $-1/2$-th power of the
size $T$. This result would follow if the central limit theorem could
be applied to the roughness as given in Eq.\ 
(\ref{eq:def-roughness-fourier}).  This is analogous to the mean field
behavior beyond the upper critical dimension of a statistical
mechanical system exhibiting a second order phase transition.

Since the $\alpha=1$ value is a threshold in the sense that
for $\alpha>1$, the scales of all cumulants diverge as
$T^{\alpha-1}$, while for  $\alpha<1$ the average becomes independent of
$T$ and higher cumulants vanish, the natural scaling for the two cases
is different.  If $\alpha>1$ the scaled quantity 
\begin{equation}
  \label{eq:scaled-w-1}
  x=\frac{w_2}{\kappa_1}
\end{equation}
has, in the limit $N\to\infty$, a convergent PDF, devoid of any
adjustable parameter.  In the following, we shall refer to the use of
variable $x$ as scaling by the average.  For $\alpha\le 1$ the same
PDF would be a Dirac delta centered on $1$, so one rather resorts to a
scaling that effectively widens the delta peak.  One can do that by
introducing \cite{Holdsworth1,Holdsworth2}
\begin{equation}
  \label{eq:scaled-w-2}
  y=\frac{w_2 - \kappa_1}{\sqrt{ \kappa_2 }},
\end{equation}
that is, scaling the roughness by the variance.  The PDF as a function
of $y$ will clearly become Gaussian below $\alpha=1/2$, but develop a
nontrivial shape for $1/2<\alpha$.  This variable was used in our
recent Letter on the $\alpha=1$ case \cite{us}, where we used the
notation $x$ for it.  We will adopt the notation that functions of $x$
and $y$ carry the labels $1$ and $2$, respectively. So, e.g., the PDF
of the scaled quantity (\ref{eq:scaled-w-1}) will be denoted by
$\Phi_{1}(x)$. 

The two scalings (\ref{eq:scaled-w-1}) and (\ref{eq:scaled-w-2}) can
lead to dramatically different shapes.  For instance, for $\alpha\to
1$ from above, Eq. (\ref{eq:scaled-w-1}) yields a peak centered at
$x=1$ and leads to a Dirac delta  at $x=1$ for $\alpha<1$.
However, (\ref{eq:scaled-w-2}) gives smooth functions continuously changing
when $\alpha$ passes through $1$ and accordingly,  we will
use sometimes (\ref{eq:scaled-w-2}) in the region $\alpha>1$ for the
sake of comparison.

\section{Scaling function for periodic boundary condition}
\label{sec:scaling-fcn-pbc}

\subsection{Analytic approach for general $\alpha$}  
\label{sec:gen-pbc}

\begin{figure}[htb]
\includegraphics{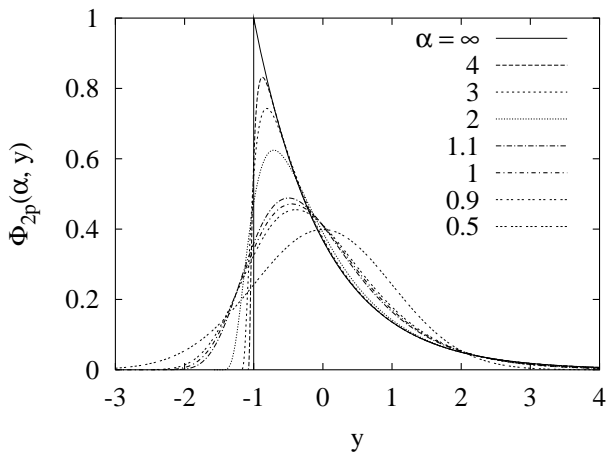}
\caption{Roughness distribution of periodic signals, Eq.\
  (\ref{eq:P_y}), for different values of $\alpha$ as a
      function of the scaling variable (\ref{eq:scaled-w-2}). Note
      that the PDF is a smooth function of $\alpha$ for all $\alpha$-s.}
\label{fig:ap0}
\end{figure}

\begin{figure}[htb]
\includegraphics{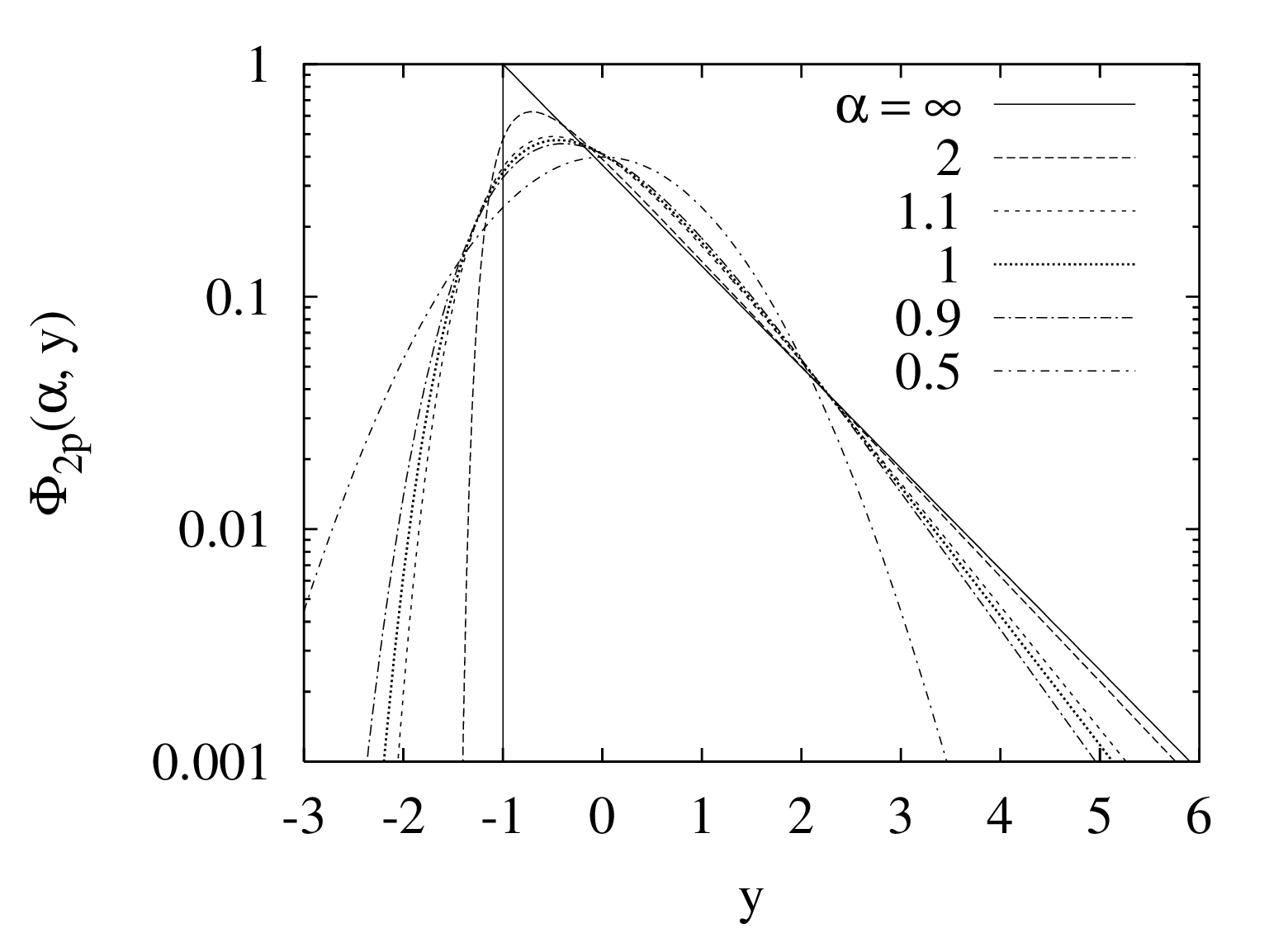}
\caption{The same as Fig.~\ref{fig:ap0} but on log-linear scale.}
\label{fig:ap0_log} 
\end{figure}

Knowing the explicit form of the generating function
(\ref{eq:gen-fun-roughness}), the roughness distribution for any
finite $N$ can be obtained by inverting the Laplace
transformation (\ref{eq:gen-fun})
\begin{equation}
P_{\textrm{p}}(\alpha, w_2, T) = \int\limits_{-i\infty}^{i\infty}
\frac{ds}{2 \pi i}  
\,e^{w_2 s} \prod_{n=1}^{N} 
\left(1 + \frac{s}{\sigma T^{1-\alpha}n^\alpha} \right)
\label{eq:pdf-N-finite}
\end{equation}
From here we concentrate on the $N \to \infty$ limit. In this limit 
the PDF of the scaling variable $x$ or $y$, defined in Eqs.\  
(\ref{eq:scaled-w-1},\ref{eq:scaled-w-2}), becomes independent of $T$.
Scaling by the variance, that is, using the $y$ variable of Eq.\  
(\ref{eq:scaled-w-2}), has the advantage to yield smooth PDF-s for all
$\alpha$-s.  We refer to this type of scaling by the label $2$ on the
PDF.  For $\alpha >1/2$ that PDF is 
\begin{eqnarray} 
\Phi_{2\textrm{p}}(\alpha, y) &=&
\sqrt{\kappa_2} P_{\textrm{p}}(\alpha, w_2, T)  \nonumber \\
=\sqrt{\zeta(2\alpha)}\hspace{-15pt}&&
\int\limits_{-i\infty}^{i\infty} \frac{ds}{2 \pi i} 
\,e^{\sqrt{\zeta(2\alpha)} y s} \prod_{n=1}^{\infty} 
\frac{ e^{\frac{s}{n^\alpha}}} {1+{\frac{s}{n^\alpha}}}
\label{Pwsc} ~.
\end{eqnarray}
The integrand has simple poles along the real axes 
and their contributions can be easily collected
\begin{equation}
\Phi_{2\textrm{p}}(\alpha, y) = \sqrt{\zeta(2\alpha)}
\sum_{m=1}^\infty m^\alpha  e^{-m^\alpha \sqrt{\zeta(2\alpha)} y-1}
Y_2(\alpha, m) ~, 
\label{eq:P_y}
\end{equation}
where 
\begin{equation}
Y_2(\alpha, m) = \prod_{n=1,\ne m}^\infty 
\frac{e^{-\left(\frac{m}{n}\right)^\alpha}}
{1-\left(\frac{m}{n}\right)^\alpha} ~.
\label{eq:Y} 
\end{equation}  
This series can be considered as the large-$y$ expansion of the PDF.
It is a general formula for any $\alpha>1/2$ and can be evaluated
numerically.  First, the precise numerical values of $Y_2(\alpha, m)$
should be determined up to a certain $m_{\textrm{max}}$. Greater
$m_{\textrm{max}}$ is needed for smaller values of $y$.  Since the
sign of $Y_2(\alpha, m)$ is known to be $(-1)^{m-1}$, it suffices to
evaluate $\left| Y_2\right|$.  The logarithm of $\left| Y_2\right|$
can be written as an infinite sum, which can be tackled numerically.
Once the values of $Y_2(\alpha, m)$ are given, the summation in
(\ref{eq:P_y}) can be done easily due to the exponential cutoff in
$m$.  The resulting PDF is depicted on Figs.\ \ref{fig:ap0} and
\ref{fig:ap0_log} for several values of $\alpha$. On Fig.\ 
\ref{fig:ap0_log} logarithmic scale was used for the ordinate to make
the tails visible for several decades.  It is clearly seen on these
figures that the shape of the PDF depends smoothly on the value of
$\alpha$. 

\begin{figure}[htb]
\includegraphics{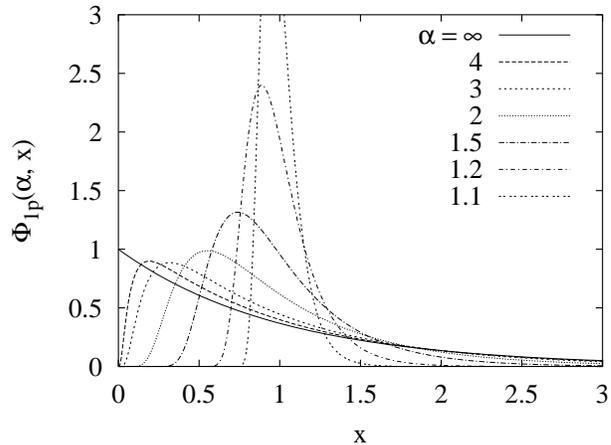}
\caption{ \label{fig:ap1} Roughness distribution of periodic signals
  for different values of $\alpha$ as a function of the scaling
  variable (\ref{eq:scaled-w-1}). Note that as $\alpha \to 1$ the PDF
  dramatically sharpens and converges to a Dirac delta function
  centered at $x=1$.}
\end{figure}

When $\alpha>1$, the natural choice for scaling variable is $x$ of
Eq.\ (\ref{eq:scaled-w-1}), and the PDF in terms of $x$ can be obtained by
changing the variable in (\ref{eq:P_y}) and making some
simplifications
 
\begin{eqnarray}
\Phi_{1\textrm{p}}(\alpha, x) &=&
\kappa_1 \Phi_{\textrm{p}}(\alpha, w_2, T)  \nonumber \\
&=& \zeta(\alpha) \sum_{m=1}^\infty m^\alpha 
e^{- x m^\alpha \zeta(\alpha)} Y_1(\alpha, m)
\label{eq:P_x} 
\end{eqnarray} 
with  
\begin{equation}  
\label{eq:prod_x}  
Y_1(\alpha, m) = \prod_{n=1,\ne m}^\infty 
\left[1-\left(\frac{m}{n}\right)^\alpha \right]^{-1} ~.
\end{equation} 
Note that it is more appropriate for numerical evaluation to use Eqs.\ 
(\ref{eq:P_y},\ref{eq:Y}) and then change the variable $y$ to $x$.
The PDF-s on Fig.\ \ref{fig:ap1} were calculated this way.  One can
observe the singularity at $x=1$ emerging as $\alpha$ approaches 1.
 
For integer values of $\alpha>1$ the infinite product
(\ref{eq:prod_x}) can be written in a closed form
\begin{equation}
Y_1(\alpha, m) = \frac{m!(-1)^{m-1}}{\alpha} 
\prod_{k=1}^{\alpha-1} \Gamma(1-a^k m) ~,
\label{eq:X}
\end{equation}
where $a=\exp(2\pi i/\alpha)$.  
For $\alpha=2$  we recover the random walk result of Ref.~\cite{FORWZ}
\begin{equation}
Y_1(2,m) = 2 (-1)^{m-1} ~.
\end{equation}
In the $\alpha=4$ case there are three factors in the product and  using
some basic properties of the $\Gamma$ function one obtains
\begin{equation}
Y_1(4,m) = \frac{4\pi m (-1)^{m-1}}{\sinh(\pi m)} ~,
\end{equation}
a result which agrees with that of Ref.~\cite{PRZ94}.

Using (\ref{eq:X}), we can write the roughness distribution for any
integer $\alpha>1$ in the relatively simple form
\begin{eqnarray}
\Phi_{1\textrm{p}}(\alpha, x) & =&
\alpha \zeta(\alpha) \sum_{m=1}^\infty
\frac{(-1)^{m-1} m^\alpha} {m!} \nonumber \\ && \times 
e^{-m^\alpha \zeta(\alpha) x}  
\prod_{k=1}^{\alpha-1} \Gamma(1-a^k m).
\end{eqnarray} 
The curves for PBC in all figures with integer $\alpha>1$ were drawn
based on this formula.

\subsection{ Special cases} 
\label{ssec:pbc-sc}

\subsubsection{$1/f$ noise ($\alpha=1$)}
\label{sssec:pbc-alpha-1}

Here we briefly revisit the case of $1/f$ noise presented in our
recent Letter \cite{us}. The natural scaling goes now by the variance
as in (\ref{eq:scaled-w-2}).  From Eq.\ (\ref{Pwsc}) the generator of
the scaling function $\Phi_{1\textrm{p}}(1,x)$ is
\begin{equation} 
  \label{eq:gen-fun-1f} 
  G_{2\textrm{p}}(1,s) = \prod_{n=1}^\infty
  \frac{e^{s/an}}{1+\frac{s}{an}},  
\end{equation} 
where the upper limit of the product could safely be taken to infinity
and $a=\pi/\sqrt{6}$.  This formula produces the Gamma function
as
\begin{eqnarray} 
  G_{2\textrm{p}}(1,s) &=& e^{\gamma s/a}\, \Gamma\left(
  1+\frac{s}{a}\right) \nonumber \\ &=& \int_0^\infty du\, e^{-u}\, \left(
  u\,e^\gamma\right)^{s/a}, \label{eq:gamma-fcn} 
\end{eqnarray} 
where $\gamma$ is Euler's constant and we also displayed Euler's
integral formula for the Gamma function \cite{as}.  Introducing the
variable $x=-(\ln u+\gamma)/a$ we finally get
\begin{subequations}  
  \label{eq:f1} 
\begin{eqnarray} 
  \label{eq:f11} 
 G_{2\textrm{p}}(1,s) &= & \int_{-\infty}^\infty dx\, e^{-sx}
  \Phi_{2\textrm{p}}(1,x), \\
  \label{eq:ftg} 
  \Phi_{2\textrm{p}}(1,x) &= & a\, e^{-(ax+\gamma)-e^{-(ax+\gamma)}}.  
 \end{eqnarray} 
\end{subequations}
The inverse Laplace transformation on (\ref{eq:f11}) gives
(\ref{eq:ftg}), so $\Phi_{2\textrm{p}}(1,x)$ is the sought PDF of the
roughness scaled by the variance.  Note that strictly speaking
(\ref{eq:f11}) is not a Laplace transformation anymore, due to the
shift of the average of $\Phi_{2\textrm{p}}(1,x)$ to zero, but the
inverse transformation can still be performed.  In such cases the
Fourier transformation is better suited for the generating function,
because the variable $y$ of Eq.\ (\ref{eq:scaled-w-2}) is not
restricted to non-negative numbers, but we could still derive the
scaling function (\ref{eq:ftg}) within the Laplace formalism.
 
The formula (\ref{eq:ftg}) is a special case of what is known as the
Fisher-Tippet{\-}{-Gumbel} function that emerges in extreme value
statistics \cite{Gumbel,Galambos}.  This comes about in a nutshell as
follows.  Suppose we have a random variable with some generic PDF, and
we draw $M$ times independently from this distribution.  The PDF of
the $k$-th largest of all those values, that is, the extreme value
PDF, will be centered around a median increasing with $M$.  Obviously,
for $M\to\infty$ and $k$ fixed, the extreme value PDF will be
determined by the tail of the original PDF.  In this limit, an
appropriate linear rescaling of the $k$-th largest value yields
generally an $M$-independent scaling function for the extreme value
PDF.  Now, in practical terms, if the original PDF has a tail decaying
faster than any power law, then the scaling function is the
Fisher-Tippet-Gumbel distribution.  Our scaling function
(\ref{eq:ftg}) corresponds to the $k=1$ case, scaled to have zero
average and unit variance.
 
The fact that (\ref{eq:ftg}) is related to extremal value statistics
does not reveal automatically the mechanism of selection of extremes
in the case of the roughness of a signal.  Also, it should be
emphasized that we do not have the usual extremal value distribution
functions for $\alpha$-s other than one.  So our result raises the
problems of why the $1/f$ noise is distinguished among all $\alpha$-s
and how extreme value selection comes about there rather than resolves
them. 
 
\subsubsection{Wiener process ($\alpha=2$) } 
\label{sssec:pbc-alpha-2} 
 
For the Wiener process with PBC the generating function as well as the
asymptotics of the PDF for small and large $x$ has been derived in
\cite{FORWZ}.  Interestingly, the distribution can be expressed in
terms of a known function.  The normalization (\ref{eq:scaled-w-1}),
natural for $\alpha>1$, will be used.  For the integrated density with
$\alpha=2$ we get from (\ref{eq:P_x})
\begin{eqnarray}
  M_{1\textrm{p}}(2,x) &=& \int_0^x \Phi_{1\textrm{p}}(2,\bar{x})\,
  d\bar{x}  \nonumber
  \\    &=& 1 + 2 \sum_{k=1}^{\infty} (-1)^k  e^{-\pi^2 k^2 x /6}\nonumber
  \\    &=&  \vartheta_4(0,e^{-\pi^2 x /6}).  
  \label{eq:int-dens}
\end{eqnarray}
$\vartheta_4$ is Jacobi's 4th Theta function \cite{as} and its
second argument is now the relevant variable \cite{jacobi}.
 
The PDF can also be written in a closed, implicit form in terms of complete
elliptic integrals of the first and second kind ($K(k)$ and $E(k)$):
\begin{subequations}
\begin{eqnarray}
\Phi_{1\textrm{p}}(2,x(k)) &=& \frac{1}{3\sqrt{2\pi}}
    (1-k^2)^{1/4}\nonumber \\ 
    && \times \,  K^{3/2}(k) 
\left[ K(k) -E(k) \right] \\
x(k) &=& \frac{6 K(\sqrt{1-k^2})}{\pi K(k)} ~. 
\end{eqnarray}
\end{subequations}

The sum over poles described in section \ref{sec:gen-pbc} for a
general $\alpha$ can be understood as the large-$x$ expansion of the
distribution.  On the other hand, we also have a small-$x$ series for
the special $\alpha=2$ case.  This can be constructed by scaling the
generating function (\ref{eq:gen-fun-roughness}) according to Eq.\ 
(\ref{eq:scaled-w-1}),  expanding it as
\begin{eqnarray}
  \label{eq:sinh-expand}
   G_{1\textrm{p}}(2,s)&=&   \frac{\sqrt{6s}}{\sinh\sqrt{6s} }
   \nonumber \\ &=& 2 \sqrt{6s} \sum_{k=0}^\infty e^{-(2 k+1)
   \sqrt{6s}},  
\end{eqnarray}
and using the inverse Laplace transform
\begin{equation}
  \label{eq:laplace-of-one-term}
  \int_{c-i\infty}^{c+i\infty} \frac{ds}{2\pi i} \sqrt{s}\, e^{-a
  \sqrt{s}} = \frac{a^2 - 2x}{4\sqrt{\pi x^5}} e^{-a^2/4x}.   
\end{equation}
The result is 
\begin{eqnarray}
  \Phi_{1\textrm{p}}(2,x)  &=&
  \int_{c-i\infty}^{c+i\infty} \frac{ds}{2\pi
  i}\,e^{sx}\,G_{1\textrm{p}}(2,s) \nonumber   \\ &&
  \hspace{-50pt} = 
  \sqrt{\frac{6}{\pi x^5}} \sum_{k=0}^\infty \left[  
  3(2k+1)^2-x\right] e^{-3(2k+1)^2/2x}.  
  \label{eq:laplace-of-sum}
\end{eqnarray}
One recovers here the nonanalytic small-$x$ asymptotics shown in
\cite{FORWZ}, and one also has the corrections to it.  Three terms
from the sum suffice to produce the PDF $\Phi_{1\textrm{p}}(2,x)$ with a
uniform error bound of $\epsilon=4 \times 10^{-5}$ in the sense that
either the approximation is within $\epsilon$ from the PDF, or the PDF
is less than $\epsilon$.  The precision of this approximation is,
however, much better when the PDF is of order unity, to give a feeling
of it one gets $29$ digits of the PDF at $x=1$.  The series can be
considered as the small-$x$ expansion of the derivative of the Theta
function (\ref{eq:int-dens}), not available in \cite{as}.  The
formulas of Sections \ref{sssec:pbc-alpha-1} and
\ref{sssec:pbc-alpha-2} can help to test the numerical evaluation of
the series (\ref{eq:P_y}-\ref{eq:prod_x}).
 
\subsubsection{ Large $\alpha$ limit} 
\label{sssec:pbc-alpha-infty} 
 
For large $\alpha$ the lowest frequency mode dominates.  Indeed, in
the action (\ref{eq:def-action}) the coefficients $c_n$ with $n>1$
will have very small variance compared to the $n=1$ case, so
practically they are zero.  The corresponding generator function is
obtained from (\ref{eq:gen-fun-roughness}) by omitting all $n>1$
factors.  Applying the scaling (\ref{eq:scaled-w-1}) we get
\begin{equation}
  \label{eq:gen-large-alpha}
  G_{1\textrm{p}}(\infty,s) = (1+s)^{-1},
\end{equation} 
whence inverse Laplace transformation yields
\begin{equation}
  \label{eq:pbc-pdf-large-alpha}
  \Phi_{1\textrm{p}}(\infty,x) = e^{-x}.
\end{equation}
Recently in Ref.\ \cite{Duna} a PDF was introduced and defined by
cumulants for arbitrary dimensions.  Its special case $d=1$
corresponds to our cumulants in Eq.\ (\ref{eq:cum-roughness}). There
the $\chi^2$ density with $2d$ degrees of freedom for large $\alpha$
was found, that for $d=1$ indeed gives (\ref{eq:pbc-pdf-large-alpha}).

\section{Scaling function for window boundary condition}
\label{sec:scaling-fcn-wbc}

\subsection{Simulations} 
 
Although periodic $1/f^\alpha$ signals exist
\cite{{FORWZ},{PRZ94},{us}}, most of the experimental signals
displaying $1/f^\alpha$ spectrum are not periodic. Having a long
experimental signal, however, the roughness can be calculated for
small uncorrelated segments (these are called the windows) and the
roughness distribution can be constructed for this ``window'' boundary
condition, abbreviated as WBC.  It is a plausible assumption that this
PDF does not depend on the BC-s of the original signal provided that
the size of the window, $T$, is much smaller than the size of the
entire signal $T_{\textrm{p}}$. Having uncorrelated windows would
require large distances between them, however, in our experience the
PDF remained unchanged even for overlapping windows.  In the
simulations of this chapter always non overlapping, but neighboring
windows were used.  Therefore it appears that the PDF of a Gaussian,
perfect $1/f^\alpha$ signal with WBC can be computed having a long
periodic signal of the same type.  This method was applied numerically
for general $\alpha$ and we report analytical results only for
$\alpha=2$ and $\alpha=\infty$ in Sec.~\ref{ssec:wbc-spec}.
 
The most accepted numerical way of generating a Gaussian $1/f^\alpha$
signal is to generate a Gaussian white noise first, filter the Fourier
spectrum of it in order to get the desired $1/f^\alpha$ behavior
(i.e., after a fast Fourier transformation its real and imaginary
parts are multiplied by $f^{-\alpha/2}$) and finally transform it
back. In this way the modes have amplitudes fluctuating around the
desired $1/f^\alpha$ value, and have random phases. Calculating the
roughness distribution of such periodic signals made it possible to
check our theoretical predictions for periodic signals.  The main
advantage of generating periodic $1/f^\alpha$ signals is, however,
that one can simulate WBC this way.
 
Having the desired periodic signal of length $T_{\textrm{p}}$ one can
construct easily the PDF of the roughness of non-overlapping parts of
size $T$.  The value of $T_{\textrm{p}}$ was chosen to be at least
$2^{20}$ while $T$ varied between $2^{6}$ and $2^{18}$. The PDF
converged to a size independent shape for each values of $\alpha$ we
have studied. The finite size effect was larger for smaller values of
$\alpha$. For $\alpha=1$ the PDF for $T=2^{6}$ was already within a
linewidth to the limit curve, however, for $\alpha=1/2$ such precision
was only reached for $T=2^{18}$.

\begin{figure}[htb]
  \includegraphics{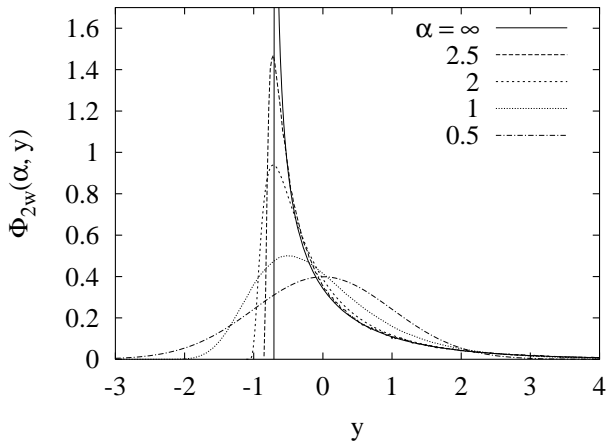}
\caption{Roughness distribution with WBC for different 
  values of $\alpha$ as a function of the scaling variable
  (\ref{eq:scaled-w-2}).  Note that the $\alpha=1/2$, 2, and $\infty$
  curves are exact results.}
\label{fig:aw0}
\end{figure}

\begin{figure}[htb]
\includegraphics{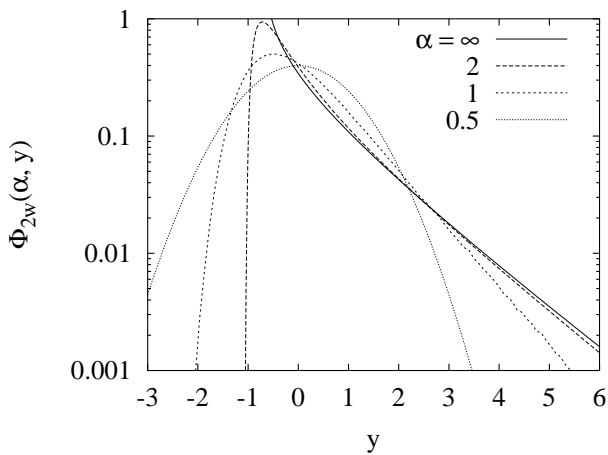} 
\caption{The same as Fig.~\ref{fig:aw0} but on log-linear scale.}
\label{fig:aw0_log}
\end{figure}

The results for a few values of alpha are depicted on Figs.\ 
\ref{fig:aw0}-\ref{fig:aw0_log} together with the two analytical
results for WBC of Sec.~\ref{ssec:wbc-spec}, which provided a good
test for the numerics.  One observes that the curves are changing
continuously with alpha in the whole range of [1/2, $\infty$]. For
$\alpha=1/2$ we recovered the Gaussian PBC result.  It is also
interesting to note that the $\alpha \to \infty$ convergence is
noticeably faster than that in the PBC case. For WBC already the
$\alpha=3$ PDF can be well approximated by that for $\alpha=\infty$.

\begin{figure}[htb]
\includegraphics{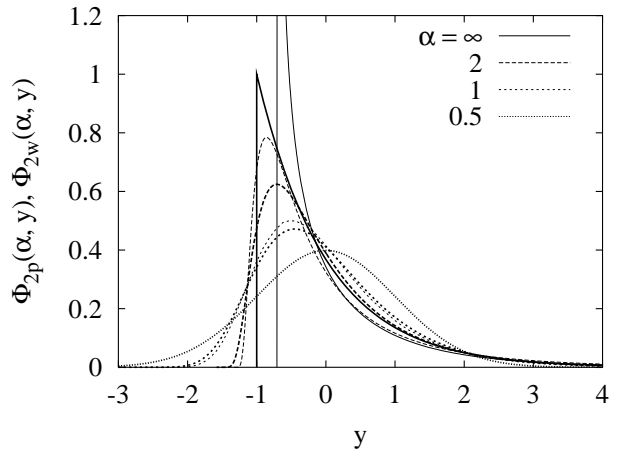} 
\caption{Comparison of PBC (thick lines) and WBC 
  (thin lines) for different values of $\alpha$ as a function of the
  scaling variable (\ref{eq:scaled-w-2}).}
\label{fig:acom0}
\end{figure}

On Fig.\ \ref{fig:acom0} PDF-s with different BC-s are compared for
several values of $\alpha$. As we have already mentioned before, for
$\alpha=1/2$ both BC result in a Gaussian PDF.  For $\alpha=1$, as we
already reported in Ref.~\cite{us}, the difference between the PDF-s
are relatively small, namely it was comparable to the precision of the
experimental values.  On Fig.\ \ref{fig:acom0} one can observe that
the difference between the PDF with PBC and WBC are larger for larger
values of $\alpha$. 

We should emphasize that the thresholds $\alpha=1$ and $\alpha=1/2$
were found to play the same role for PBC and WBC.  Namely, the natural
scaling of the roughness goes by the average and by the variance for
$\alpha>1$ and $1>\alpha$, respectively, and for $\alpha\le 1/2$ the
scaled PDF becomes a Gaussian for both BC-s.  This evidence strongly
supports our expectation that the scaling of cumulants for PBC, as
summarized in Table \ref{tab:scaling}, also applies to WBC.

It is worth recalling that critical exponents in second order phase
transitions are generally believed not to depend on BC, but the
scaling functions do vary below the upper critical dimension, the
threshold for mean field behavior, for different BC-s
\cite{Wu,exotic,Wu2}.  The aforementioned BC-independence of the scaling
of the cumulants on the one hand, and the BC-dependence of the scaled
PDF for the roughness on the other one, show a close analogy with
critical phenomena.  The threshold for mean-field-type behavior is now
$\alpha=1/2$, because below that the PDF is Gaussian.

\subsection{Analytic results in special cases}  
\label{ssec:wbc-spec}
\subsubsection{Wiener process ($\alpha=2$)}
\label{sssec:wbc-alpha-2}

Since the increments of the Wiener process are uncorrelated, the
trajectory in a window is coupled to the path outside only by its two
endpoints. However, if the length of the outer trajectory is much
larger than that of the window then the endpoints appear essentially
as unconstrained on the scale of the inner trajectory.  Thus the WBC
corresponds to the free BC in the $\alpha=2$ case.  The generating
function for free BC is, apart from scaling of $s$, the square root of
the generating function for PBC as given below
\begin{equation}
  \label{eq:wbc-gen-fun}
    G_{1\textrm{w}}(2,s) =  \frac{\sqrt[4]{12 s}}{\sqrt{\sinh\sqrt{12
    s}}},   
\end{equation}
where the normalization
${G_{1\textrm{w}}}^\prime(2,s)|_{s=0}=-1$ was used,
corresponding to scaling by the average (\ref{eq:scaled-w-1}).  This
yields the PDF $\Phi_{1\textrm{w}}(2,x)$ by inverse Laplace
transformation.  Eq.\ (\ref{eq:wbc-gen-fun}) can be understood if one
notes that free BC means that the action is diagonalized by cosine
eigenfunctions, having zero derivatives at the endpoints.  So the
spectrum is labeled by integers, like for PBC, but there is no
degeneracy. The generating function is in essence the reciprocal
square root determinant.  While for PBC the square root disappears due
to the two-fold degeneracy, in free BC the root remains.  A more
rigorous derivation will be presented in \cite{gr}.

The small-$x$ series of the PDF ${\Phi_{1\textrm{w}}}(2,x)$
can be obtained by expanding Eq.\ ({\ref{eq:wbc-gen-fun}) as
\begin{equation} 
  \label{eq:wbc-root-sinh-expand}
  G_{1\textrm{w}}(2,s) = 2 \sqrt[4]{3 s}
  \sum_{k=0}^{\infty} (-1)^k \tbinom {-1/2}{k} e^{-(4k+1)\sqrt{3
  s}}, 
\end{equation} 
and using the Laplace transform of a single term in the sum as
\begin{eqnarray}
  \label{eq:wbc-laplace-of-one-term}
  \int_{c-i\infty}^{c+i\infty} \hspace{-18pt}&& \frac{ds}{2\pi i}
  \sqrt[4]{s}\, e^{-a \sqrt{s}}  \nonumber \\
  &=&\frac{a^{3/2}}{2\sqrt{2\pi} x^2} e^{-a^2/4x} ~_2F_0
  \left(-\frac{1}{4}, -\frac{3}{4};-\frac{4x}{a^2}\right),    
\end{eqnarray}
where $_2F_0$ is a hypergeometric function in standard notation
\cite{as}.  Based on these relations we finally get
\begin{eqnarray}
    \label{eq:pfbc} 
  \Phi_{1\textrm{w}}(2,x) &=& \int_{c-i\infty}^{c+i\infty} \frac{ds}{2\pi i}
  \, e^{sx}\,G_{1\textrm{w}}(2,s) \nonumber \\  
   & =& \frac {3} {x^2\sqrt{2\pi}} \,
 \sum_{k=0}^{\infty}\, (4k+1)^{3/2} \frac{(2k-1)!!}{2^{k}\, k!} 
 \nonumber \\ && \times\, \exp\left( -\frac{3(4k+1)^2}{4x}\right)  \nonumber\\
  && \times\,\,\,   _2F_0\left(-\frac{1}{4}, -\frac{3}{4};
  -\frac{4x}{3(4k+1)^2}   \right).   
\end{eqnarray} 
Although this can be considered as the small-$x$ expansion, it is
enough to retain the first three terms to get an estimation of
$\Phi_{1\textrm{w}}(2,x)$ to a uniform precision of $\epsilon=6
\times 10^{-6}$ for any $x$.  Similarly to the PBC case, this is
understood such that either the approximation deviates from the PDF by
less than $\epsilon$, or, the PDF is smaller than $\epsilon$.  To
illustrate the accuracy of three terms from (\ref{eq:pfbc}) for
reasonable $x$-s, at $x=1$ they happen to give $53$ digits of the PDF
correctly.  Series (\ref{eq:pfbc}) has been used to draw the
$\alpha=2$ curve of Figs.\ \ref{fig:aw0}, \ref{fig:aw0_log},
\ref{fig:acom0}.

\subsubsection{Large $\alpha$ limit} 
\label{sssec:wbc-alpha-infty}

As it has been discussed in Sec.\ \ref{sssec:pbc-alpha-infty}, for
large $\alpha$ only the lowest frequency mode dominates the
trajectory.  In the full time interval with PBC, $T_{\textrm{p}}$, this means
that
\begin{equation}
  \label{eq:wbc-single-mode}
  h(t)= r \cos\left( \frac{2\pi t}{T_{\textrm{p}}} + \varphi\right),
\end{equation}
where $ c_1= re^{i\varphi}/2$ is the Fourier coefficient of the $n=1$
mode. The fact that the real and imaginary parts of $c_1$ are
independent, identically distributed, Gaussian variables has the
consequence that the PDF for the polar parameters is
\begin{equation}
  \label{eq:wbc-pdf-r-phi}
  \rho(r,\varphi) = \frac{r}{2\pi a} e^{-r^2/2a},
\end{equation}
where we do not specify the variance $a$ as it will disappear from
the final formula anyhow. 
 
Now we consider within the overall periodic signal $h(t)$ a small
window of length $T \ll T_{\textrm{p}}$.  Since we shall average over the phase,
we can take the window to be $[0,T]$.  An expansion of
(\ref{eq:wbc-single-mode}) up to $t^2$ shows that the roughness of the
trajectory within the window is in leading order
\begin{equation}
  \label{eq:wbc-roughness}
  w_2 =  r^2 b \sin^2\varphi, 
\end{equation} 
where $b$ is a constant depending on $T$ and $T_{\textrm{p}}$, whose value will
turn out to be immaterial.  We thus have for the PDF of the roughness
\begin{eqnarray} 
  P_{1\textrm{w}}(\infty,w_2) &=& \int_0^\infty dr \int_0^{2\pi}
  d\varphi\, \rho(r,\varphi) \nonumber \\
  && \times  \delta\left(w_2-r^2 b
  \sin^2\varphi\right). 
  \label{eq:wbc-pdf-roughness}
\end{eqnarray}
Inserting expression (\ref{eq:wbc-pdf-r-phi}) for $\rho(r,\varphi)$ we
get for the scaled variable $x=w_2/\left<w_2\right>$ the PDF
\begin{equation}
  \label{eq:wbc-pdf}
   \Phi_{1\textrm{w}}(\infty,x) = \frac{e^{-x/2}}{\sqrt{2\pi x}}.
\end{equation}

Note that, similarly to the Wiener process, the large $\alpha$ limit
of the generating function for the WBC is essentially the square root
of that for PBC, apart from a scale change due to normalization.
Indeed, the generating function for PBC is given by
(\ref{eq:gen-large-alpha}), while the Laplace transform for the above
PDF is $(1+2s)^{-1/2}$.

\section{$1/f^\alpha$ surfaces in arbitrary dimensions}
\label{sec:gen-dim}

So far we have considered random fields of one component that were
functions of time, that is, a $1+1$ dimensional system.  A natural
generalization is to consider $d+1$ dimensional random surfaces, where
the substrate has $d$-dimensional coordinates $\bm{x}$, and the field
$h(\bm{x})$ still has one component.  Imposing PBC now means that the
substrate is a $d$-dimensional torus.  Again, we define the
probability density functional of a surface through an action that
depends on the Fourier components $c_{\bm{n}}$ of the surface,
$\bm{n}=(n_1,\dots n_d)$, $n_j=-N,\dots N$ being integers, and
$c_{\bm{n}}=c_{-\bm{n}}^\ast$.  The spatial unit is $\ell$, the length
of a period $L=N\ell$, the unit volume $v=\ell^d$, and the total
$d$-dimensional volume $V=L^d$.  (In the case of a usual surface $d=2$
and $V$ is the area of the substrate.)  The action is
\begin{equation}
  \label{eq:action-gen-dim}
  S\left( \left\{ c_{\bm{n}} \right\} \right) = 2\sigma
  V^{1-\alpha/d} {\sum_{\bm{n}}}^\prime \left| \bm{n}\right| ^\alpha
  \left|c_{\bm{n}}\right|^2. 
\end{equation}
Here prime means that the summation excludes the origin and counts
only half of the remaining index space so that if a vector $\bm{n}$ is
included then $-\bm{n}$ is not.  For $\alpha=2$ this is the action
associated with the stationary distribution of the Edwards-Wilkinson
model \cite{surface-review}.  In the case of general $\alpha$-s, Eq.\ 
(\ref{eq:action-gen-dim}) is the long-range interaction part of the
single-component version of the generalized $O(n)$ Hamiltonian of
\cite{FMN}, for a recent reference see \cite{Cham01}.  We shall
briefly review some scaling properties for arbitrary dimensions in
order to put $1/f^\alpha$ noise, as discussed in previous Chapters, in
a broader perspective.

The roughness of a surface is a random variable, whose PDF can be
derived similarly to the $d=1$ case discussed in Ch.\ 
\ref{ssec:roughness}. The short-range interaction case, $\alpha=2$,
has been studied in detail by Refs.\ \cite{RP94, Holdsworth1,
  Holdsworth2} for $d=2$ and by \cite{Bram01} for arbitrary dimension.
The generating function of the PDF $P(w_2)$ is obtained as
\begin{equation}
  \label{eq:fun-roughness-gen-dim}
  G(s) = {\prod_{\bm n}}^\prime \left( 1+\frac{s}{\sigma
  V^{1-\alpha/d} \left| \bm{n} \right| ^\alpha}
  \right)^{-1},  
\end{equation}
whence the cumulants are 
\begin{equation}
  \label{eq:cum-roughness-gen-dim}
  \kappa_k = \frac{(k-1)!}{\left(\sigma V^{1-\alpha/d}\right)^k}
  {\sum_{\bm n}}^\prime \frac{1}{\left| {\bm n}\right| ^{ \alpha k}}.   
\end{equation}
Note that the cumulants derived here were used to define the model of
Ref.\ \cite{Duna}.  The sum converges for $\alpha k>d$, diverges
logarithmically as $\ln N\propto \ln(V/v)$ for $\alpha k=d$ and like a
power function as $N^{d-\alpha k}\propto (V/v)^{1-\alpha k/d}$ for
$\alpha k<d$.  The scaling properties of the cumulants are summarized
in Table \ref{tab:scaling-gen-dim}.

\begin{table}
\caption{\label{tab:scaling-gen-dim} Scaling of the cumulants for various
  $\alpha$-s for general dimension  The dots indicate that the last
  formula is valid from then on, i.e., ``$\dots$'' extends the validity to
  higher $k$-s and vertical dots to higher $d$-s. } 
\begin{ruledtabular}
\begin{tabular}{ccccc}
Range or& & & & \\
value of $d$.  &$\kappa_1$ &$\sqrt{\kappa_2}$
&$\sqrt[3]{\kappa_3}$ &$\sqrt[4]{\kappa_4}$
\\ 
\hline
$(0, \alpha) $ & $V^{\alpha/d-1}$ & $\dots$ & & \\
$\alpha$ & $\ln\frac{V}{v}$ & O($1$) &$\dots$ &\\
$(\alpha,2 \alpha)$ & $v^{\alpha/d-1}$ & $V^{\alpha/d-1}$ &$\dots$ &\\
$2\alpha$ & $\vdots$ & $V^{-1/2}\sqrt{\ln\frac{V}{v}}$
& $V^{-1/2}$ &$\dots$ \\
$ (2\alpha,3\alpha)$ &  &
$v^{\alpha/d-1/2}\, V^{-1/2} $ &$V^{\alpha/d-1}$ &$\dots$ \\ 
$3\alpha$ &  &
$\vdots $ &
$V^{-2/3}\sqrt[3]{\ln\frac{V}{v}}$ & $V^{-2/3}$ \\ 
$(3\alpha,4\alpha)$ & 
&  & $v^{\alpha/d-1/3}\,V^{-2/3}$ &$V^{\alpha/d-1}$\\ 
 & & &$\vdots$&\\ 
\end{tabular}  
\end{ruledtabular}   
\end{table}

For fixed $d$, in the large $\alpha$ limit, the modes with $\left|
  \bm{n}\right|=1$ dominate and so the roughness $w_2$ obeys the
$\chi^2$ distribution, as observed in Ref.\ \cite{Duna}.  For finite
$\alpha$, the threshold dimension where the mean logarithmically
diverges is $d=\alpha$.  For $d>\alpha$, however, $P(w_2)$ becomes a
Dirac delta, but if one looks at it on the scale of the variance for
large but finite $N$ then for $d<2\alpha$ a nontrivial function
emerges.  At $d=2\alpha$ the scale of the variance becomes larger than
the scale of all higher cumulants, and thus for $d>2\alpha$ the
scaling function is Gaussian.  This represents normal finite size
scaling with fluctuations of the order of $V^{-1/2}$.  Table
\ref{tab:scaling-gen-dim} is in accordance with the known fact
that $d=\alpha$ and $d=2\alpha$ can be viewed as the lower and upper
critical dimension of the system, respectively \cite{FMN}.

It is reasonable to assume that the scaling of the cumulants as
described above is not specific to the PBC used here, but generally
characterizes $d+1$ dimensional Gaussian surfaces with dispersion
exponent $\alpha$ for any BC-s.  Note that here only the scaling of
averages were considered, the evaluation of distribution functions in
arbitrary dimensions is beyond the scope of the present paper.

\section{Final remarks}
\label{sec:final}

As we have shown, the roughness distribution of periodic Gaussian
$1/f^\alpha$ signals can be calculated for arbitrary $\alpha$.  The
final expression is simple enough that it can be easily handled
numerically and the scaling functions can be displayed in the relevant
range of their argument.  Also for WBC we provide a simple method to
generate the scaling function by numerical simulation.  Examining
these scaling functions, we found an important feature in their
$\alpha$ dependence.  Namely, the shape of the functions varies
noticeably with alpha in the physically rather interesting range of
$1\le \alpha\le 2$.  This observation underlies the usefulness and
effectiveness of the roughness distribution as a tool for establishing
common or distinct origins of scale-invariant behavior in different
systems.
  
The present gallery of scaling functions is ready to be applied for
determining accurate values of $\alpha$ in scale invariant systems
where the fluctuations are Gaussian. Since we have both the PBC and
WBC scaling functions one can investigate models where PBC is used
preferentially as well as experimental systems where WBC is usually
obtained. Furthermore, the gallery can also be helpful in establishing
the presence of non-gaussian effects.  It should be clear, however, the
non-gaussian effects are on the unfinished end of our study of
roughness distributions. One can investigate the non-gaussian effects
in a given system by simulations \cite{{RP94},{KPZ}} but the real
question one should answer here is this: Can one find a classification
of nonlinear theories which produce a given $\alpha$, and can one find
the roughness distributions for the various classes? Judging from the
perspective of a related topic of critical dynamics this appears to be
a highly nontrivial question.

\begin{acknowledgments}
  Thanks are due to L. B. Kish, Z. Gingl, and P. Holdsworth for
  illuminating discussions.  This work was supported the by the
  Hungarian Academy of Sciences (Grant No. OTKA T029792) and partly by
  the Swiss National Science Foundation.
\end{acknowledgments}


\end{document}